\documentclass[11pt,a4paper]{article}

\usepackage{amsmath,amssymb}     
\usepackage{color}
\usepackage{graphicx}
\usepackage{subfigure}
\usepackage{cite}                
\usepackage{hyperref}            
\usepackage{multirow,makecell}
\usepackage{braket,bm}
\usepackage{cancel}
\RequirePackage{doi}
\usepackage{hyperref}
\usepackage{leftidx}
\usepackage[text={17cm,24.5cm},centering]{geometry}
\usepackage{textcomp}
\usepackage{wasysym}

\numberwithin{equation}{section}   

\def \be {\begin{equation}}
\def \ee {\end{equation}}
\def \ba {\begin{array}}
\def \ea {\end{array}}
\def \bea{\begin{eqnarray}}
\def \eea{\end{eqnarray}}

\def \d {\delta}

\def \l {\lambda}

\def \s {\sigma}

\def \t {\tau}

\begin{document}	
\title{\textbf {Exact Overlaps in the Lieb-Liniger Model from Coordinate Bethe Ansatz}}
\author{Hui-Huang Chen\footnote{chenhh@jxnu.edu.cn}~,}
\date{}
	
\maketitle
\underline{}
\vspace{-12mm}
	
\begin{center}
	{\it
College of Physics and Communication Electronics, Jiangxi Normal University,\\ Nanchang 330022, China\\
	}
	\vspace{10mm}
\end{center}
\begin{abstract}
	 In the paper \cite{Jiang:2020sdw}, the authors developed a new method to compute the exact overlap formulas between integrable boundary states and on-shell Bethe states in integrable spin chains. This method utilizes the coordinate Bethe ansatz representation of wave functions and singularity property of the off-shell overlaps. In this paper, we use this new method to derive the formula for overlaps between the Lieb-Liniger Bethe states and the Bose-Einstein condensate (BEC) state. As a simple application this method, we  obtained the overlaps between the Lieb-Liniger eigenstates and the free particle states with pair structure.
\end{abstract}

\thispagestyle{empty}

\newpage

\tableofcontents
\section{Introduction}
In the study of non-equilibrium statistical physics and high energy physics, in particular in the homogeneous quantum quench problem \cite{essler2016quench} and one-point function in the defect $AdS/CFT$ correspondence \cite{deLeeuw:2015hxa, Buhl-Mortensen:2015gfd, deLeeuw:2018mkd}(see \cite{deLeeuw:2017cop} for a review), the exact on-shell overlap is an important quantity. In the quantum quench framework, a system is prepared in some initial state and then is allowed to evolve in time with some Hamiltonian for which the initial state is not an eigenstate. Overlaps between the initial state and the eigenstates of the model are starting point to obtain analytical results for the post-quench time evolution \cite{caux2016quench}. This quantity is also related to exact $g$-function or boundary entropy \cite{dorey2004integrable, Kormos:2010ae} in integrable quantum field theory. Recently, people find that under some interesting set-up, the three-point function in $AdS_5/CFT_4$ can be view as an exact $g$-function in some integrable quantum field theory on the string world sheet \cite{Jiang:2019zig}.
\par The first overlap formula of the on-shell Bethe states with some particular initial state was found in \cite{brockmann2014gaudin}, it has a factorized form and is similar to Gaudin formula for the norm of the Bethe state \cite{korepin1982calculation}. People found that this factorizable property still survives in a large set of states--integrable boundary states \cite{Piroli:2017sei}(or integrable states for short). In \cite{Piroli:2017sei} integrable states are defined as the states which are annihilated by all odd local conserved charges for integrable lattice systems. In this paper, we also use this definition for the integrable continuum model.
\par In \cite{de2014solution} the authors studied the Bose-Einstein condensate (BEC) to the Lieb-Liniger quench using the quench action method where the conjectured exact overlap formula for the BEC state with the Lieb-Liniger energy eigenstate is a starting point. This conjectured overlap formula was rigorously proved in \cite{brockmann2014overlaps}, using the fact that the Lieb-Liniger model quantities can be obtained from a scaling limit of the XXZ spin chain \cite{golzer1987nonlinear} where a generalized Quantum Transfer Matrix (QTM) method can be applied \cite{Piroli:2017mmz, Pozsgay:2018ixm}.
\par Although this overlap formula is rather important, a direct proof is still missing. In a very recent paper \cite{Jiang:2020sdw}, the author proposed a new method to directly calculate the overlap between on-shell Bethe states and integrable states in integrable spin chain models using the Coordinate Bethe Ansatz (CBA) formalism. This method is very simple and inspiring and can directly apply to the Lieb-Liniger model where the Bethe wave function is known. In fact, using coordinate Bethe ansatz to calculate the overlaps between integrable states and Bethe states was firstly proposed in \cite{deLeeuw:2015hxa}. Coordinate Bethe ansatz was also applied to calculating the correlation functions in non-equilibrium problems \cite{zill2015relaxation,zill2016coordinate}.
\par In this paper, we use this newly proposed method to derive the overlap for the BEC state with the Lieb-Liniger energy eigenstate, which agrees with the early result. We also obtained overlap formulas for a special kind of $N$-particle states with the Lieb-Liniger Bethe states. For $N=2$, the state is integrable and we obtained the exact overlap. For $N\geq 4$, the states are non-integrable, we only give the overlap in the large volume (length) limit by focusing on the singular part of the formal (off-shell) overlap. Overlaps of the Bethe states with non-integrable states have been studied previously \cite{bucciantini2016stationary, deLeeuw:2019sew}. Such kind formulas are of great importance since overlap formulas serve as a starting point to the study of the related quench problem. The remaining part of this paper is organized as follows: In section \ref{sec2}, we review some basic facts about the Lieb-Liniger model and fix some notations. In section \ref{sec3}, we study the overlap formula using coordinate Bethe ansatz method started with the one- and two-particle cases which serve as warm-up examples and give hints to deal with generic cases. In the next subsection, we derive the overlap formula of the BEC state and the generic parity symmetric on-shell Bethe state.  In section \ref{sec4}, we obtained some new overlap formulas using the same techniques. Finally, we conclude in section \ref{sec5} and the proof of the vanishing of the overlap of generic on-shell Bethe states with the BEC state is given in appendix \ref{appenA}.
\section{The Lieb-Liniger model}\label{sec2}
We consider one-dimensional Boson gas on a ring of circumference $L$ with $\d$-function repulsive potential and impose periodic boundary condition. This is the famous Lieb-Liniger model, whose Hamiltonian is given by
\be
H=-\frac{\hbar^2}{2m}\sum_{j=1}^N\frac{\partial^2}{\partial x_j^2}+2c\sum_{j>k}\d(x_j-x_k),
\ee
We will set $2m=\hbar=1$ in the following for simplicity.
\subsection{The Bethe equations}
We can write down the energy eigenstates of the Lieb-Liniger Hamiltonian in the coordinate space as
\be\label{wave}
\braket{\bm{x}_N|\bm{\l}_N}=\frac{c^{N/2}}{\sqrt{N!}}\sum_{\s\in S_N}A(\bm{x}_N|\s\bm{\l}_N)\exp\{i\sum_{k=1}^N\l_{\s_k}x_k\},
\ee
where
\be
A(\bm{x}_N|\s\bm{\l}_N)=\prod_{j>k}\left[1-\frac{ic~\textrm{sign}(x_j-x_k)}{\l_{\s_j}-\l_{\s_k}}\right],
\ee
and $\ket{\bm{x}_N}$ and $\ket{\bm{\l}_N}$ is the shorthand notations of $\ket{x_1,x_2,\cdots,x_N}$ and $\ket{\l_1,\l_2,\cdots,\l_N}$ respectively.
In the above expressions, we also have used the general notation
\be
\s\bm{\l}_N=\{\l_{\s_1},\l_{\s_2},\cdots,\l_{\s_N}\}.
\ee
\par The requirement of the wave function $\braket{\bm{x}_N|\bm{\l}_N}$ should be periodic in each of its arguments $x_j$ results in the Bethe equations
\be\label{Bethe}
e^{i\l_jL}=-\prod_{k=1}^{N}\frac{\l_j-\l_k+ic}{\l_j-\l_k-ic},\qquad j=1,2,\cdots,N.
\ee
The corresponding energy eigenvalue is
\be
E=\sum_{j=1}^N\l_j^2.
\ee
When the Bethe equations eq.~(\ref{Bethe}) are not satisfied, the wave function eq.~(\ref{wave}) still defines a state which we call it off-shell. On the contrary, when the Bethe equations are instead satisfied, we call the state on-shell.
\par The Lieb-Liniger model has infinite many conserved charges $Q_1,Q_2,Q_3,\cdots$. We can identify
\be
Q_1=P=-i\sum_{j=1}^N\frac{\partial}{\partial x_j},\qquad Q_2=H=-\sum_{j=1}^N\frac{\partial^2}{\partial x_j^2}+2c\sum_{j>k}\d(x_j-x_k).
\ee
The higher charges can have very complicated expressions in the coordinate space representation \cite{davies2011higher}. For example:
\be
Q_3=i\sum_{j=1}^{N}\frac{\partial^3}{\partial x_j^3}-3ic\sum_{j>k}\d(x_j-x_k)(\frac{\partial}{\partial x_j}+\frac{\partial}{\partial x_k}).
\ee
However their actions on the Bethe states $\ket{\bm{\l}_N}$ are extremely simple:
\be
Q_n\ket{\l_1,\l_2,\cdots,\l_N}=\left(\sum_{j=1}^N\l_j^n\right)\ket{\l_1,\l_2,\cdots,\l_N},\qquad n=1,2,3,\cdots.
\ee
\subsection{The parity symmetric Bethe state}
The norm of the Bethe states is given by the Gaudin formula \cite{korepin1982calculation}
\be
\braket{\bm{\l}_N|\bm{\l}_N}=\int d^N\bm{x}_N\braket{\bm{\l}_N|\bm{x}_N}\braket{\bm{x}_N|\bm{\l}_N}=c^N\prod_{j\neq k}f(\l_j,\l_k) \det G,
\ee
where $G$ is an $N\times N$ matrix whose elements are
\be
G_{jk}=\d_{jk}[L+\sum_{l=1}^N\varphi(\l_j-\l_l)]-\varphi(\l_j-\l_k),
\ee
and the function $f(\l_1,\l_2)$ is defined as
\be\label{f}
f(\l_1,\l_2)=\frac{\l_1-\l_2+ic}{\l_1-\l_2}.
\ee
The function $\varphi(\l)$ is given by
\be
\varphi(\l)=\frac{2c}{\l^2+c^2}.
\ee
\par In this paper, we will consider the overlaps of the Lieb-Liniger energy eigenstates $\ket{\bm{\l}_N}$ with both integrable and non-integrable states.
As mentioned in the introduction, a state $\ket{\Psi}$ is called an integrable state when \cite{Piroli:2017sei}
\be
Q_{2k+1}\ket{\Psi}=0,\qquad k=0,1,2,\cdots.
\ee
Integrable states can only have non-vanishing overlaps with Bethe states with parity symmetry:
\be
\{\bm{\l}_N\}=\{\l_1^+,-\l_1^+,\cdots,\l_{N/2}^+,-\l_{N/2}^+\}\equiv\{\bm{\l}_{N/2}^+,-\bm{\l}_{N/2}^+\},
\ee
where we have assumed $N$ is an even number.
\par The norm of the on-shell Bethe states having the above pair structure can be factorized further
\be
\parallel\ket{\bm{\l}_{N/2}^+,-\bm{\l}_{N/2}^+}\parallel^2=c^N\prod_{j=1}^{N/2}f(\l_j^+,-\l_j^+)f(-\l_j^+,\l_j^+)\prod_{1\leq j<k\leq N/2}[\bar{f}(\l_j^+,\l_k^+)]^2 \det G^+\det G^-,
\ee
where\footnote{The function $\bar{f}(\l_1,\l_2)$ is symmetric in our case:$\bar{f}(\l_1,\l_2)=\bar{f}(\l_2,\l_1)$.}
\be\label{ff}
\bar{f}(\l_1,\l_2)=f(\l_1,\l_2)f(\l_1,-\l_2)f(-\l_1,\l_2)f(-\l_1,-\l_2),
\ee
and $G^{\pm}$ are $\frac{N}{2}\times \frac{N}{2}$ matrices with elements
\be
G_{jk}^{\pm}=\d_{jk}[L+\sum_{l=1}^{N/2}\varphi^+(\l_j^+,\l_l^+)]-\varphi^{\pm}(\l_j^+,\l_k^+),
\ee
with
\be
\varphi^{\pm}(\l_1,\l_2)=\varphi(\l_1-\l_2)\pm \varphi(\l_1+\l_2).
\ee
\par For future use, we introduce the variables
\be\label{a}
a(\l)=e^{i\l L},\qquad a_j\equiv a(\l_j),\quad j=1,2,\cdots, N.
\ee
Thus the Bethe equations ~\ref{Bethe} can be written as
\be
a_j=\prod_{k\neq j}\frac{f(\l_j,\l_k)}{f(\l_k,\l_j)}.
\ee
\section{Overlap between the Bethe state and the BEC state}\label{sec3}
In this section, we will consider the overlap between the Bethe state and the BEC state $\ket{BEC}$, whose wave function is constant: $\braket{\bm{x}_N|BEC}=\frac{1}{L^{N/2}}$.
\subsection{One-particle states}
As a warm-up, let's first consider the one-particle states
\be
\begin{split}
\braket{BEC|\l_1}&=\int_0^Ldx_1\braket{BEC|x_1}\braket{x_1|\l_1}=\sqrt{\frac{c}{L}}\int_0^Le^{i\l_1x_1}dx_1\\
&=\sqrt{\frac{c}{L}}\frac{e^{i\l_1L}-1}{i\l_1}
=\sqrt{\frac{c}{L}}\frac{a_1-1}{i\l_1} \qquad(\l\neq0),
\end{split}
\ee
where the definition eq.~(\ref{a}) has been used. For on-shell states, which means the Bethe equation $a_1=e^{i\l_1L}=1$ is satisfied, the above overlap is non-vanishing only when $\l=0$. The integral is trivially integrated to give the result $\braket{BEC|\l_1=0}=\sqrt{cL}$.
\par We can obtain the same result using a limiting procedure:
\be
\braket{BEC|\l_1=0}=\lim_{\l_1\rightarrow 0}\sqrt{\frac{c}{L}}\frac{e^{i\l_1L}-1}{i\l_1}=\sqrt{cL}.
\ee
\subsection{Two-particle states}
We now consider the two-particle case. For the readers' convenience and to gain more experiences for computing overlap of Bethe state with multi-particle states, we work out the details.
\par Firstly, we write
\be\label{S2}
\begin{split}
&\braket{BEC|\l_1,\l_2}=\frac{c}{L\sqrt{2!}}\Big\{\int_0^Ldx_1\int_0^Ldx_2e^{i\l_1x_1+i\l_2x_2}\left[1-\frac{ic~\textrm{sign}(x_2-x_1)}{\l_{2}-\l_{1}}\right]\\
&+\int_0^Ldx_1\int_0^Ldx_2e^{i\l_2x_1+i\l_1x_2}\left[1-\frac{ic~\textrm{sign}(x_2-x_1)}{\l_{1}-\l_{2}}\right]\Big\}\\
&=\frac{c}{L\sqrt{2!}}\Big\{f(\l_1,\l_2)\int_0^Ldx_2\int_0^{x_2}dx_1e^{i\l_1x_1+i\l_2x_2}+f(\l_2,\l_1)\int_0^Ldx_1\int_0^{x_1}dx_2e^{i\l_1x_1+i\l_2x_2}\\
&+f(\l_2,\l_1)\int_0^Ldx_2\int_0^{x_2}dx_1e^{i\l_2x_1+i\l_1x_2}+f(\l_1,\l_2)\int_0^Ldx_1\int_0^{x_1}dx_2e^{i\l_2x_1+i\l_1x_2}\Big\}.
\end{split}
\ee
\par In order to evaluate the above integral, it's useful to introduce the basic integral in the region $0<x_1<x_2<L$:
\be\label{B2def}
B_2(\l_1,\l_2;a_1,a_2)=\int_0^Ldx_2\int_0^{x_2}dx_1e^{i\l_1x_1+i\l_2x_2},
\ee
where the $a$ variables dependence of the function $B_2$ is obtained only after the integral is worked out and the definition eq.~(\ref{a}) is substituted.
Other integral can be obtained by permutating the arguments of this function.
\par Note that the last two lines in the curly bracket in eq.~(\ref{S2}) give the same result. It's easy to see that this property also holds in the multi-particle case.
Thus the two-particle formal (off-shell) overlap can be written as
\be\label{two}
\begin{split}
\braket{BEC|\l_1,\l_2}=\frac{2!c}{L\sqrt{2!}}[f(\l_1,\l_2)B_2(\l_1,\l_2;a_1,a_2)+f(\l_2,\l_1)B_2(\l_2,\l_1;a_2,a_1)],
\end{split}
\ee
where
\be\label{B2int}
B_2(\l_1,\l_2;a_1,a_2)=-\frac{a_1a_2-1}{\l_1(\l_1+\l_2)}+\frac{a_2-1}{\l_1\l_2}.
\ee
It's easy to check that when the Bethe state is on-shell, $\textup{i.e.}$ when we do the following replacement in above two-particle formal overlap formula eq.~(\ref{two})
\be
a_1=\frac{f(\l_1,\l_2)}{f(\l_2,\l_1)},\quad a_2=\frac{f(\l_2,\l_1)}{f(\l_1,\l_2)},
\ee
we get identically zero.
\par To obtain non-vanishing on-shell overlaps, we need to consider the parity-invariant states with $\l_2=-\l_1$. The formal overlap eq.~(\ref{two}) has a pole at $\l_2=-\l_1$. Around this pole, we have
\be
\braket{BEC|\l_1,\l_2}\sim\frac{a_1a_2-1}{i(\l_1+\l_2)}\frac{c\sqrt{2!}}{L}\left[\frac{f(\l_1,\l_2)}{i\l_1}+\frac{f(\l_2,\l_1)}{i\l_2}\right].
\ee
Note that when the state is on-shell, we have $a_1a_2=1$. The singular part can have a proper limit. The non-vanishing on-shell overlap only comes from this singular part.
In other words, the exact parity-invariant two-particle overlap can be obtained by taking the following limiting procedure
\be
\begin{split}
\braket{BEC|\l_1,-\l_1}&=\lim_{\l_2\rightarrow -\l_1}\frac{e^{i(\l_1+\l_2)L}-1}{i(\l_1+\l_2)}\frac{c\sqrt{2!}}{L}\left[\frac{f(\l_1,\l_2)}{i\l_1}+\frac{f(\l_2,\l_1)}{i\l_2}\right]\\
&=c\sqrt{2!}\left[\frac{f(\l_1,-\l_1)}{i\l_1}+\frac{f(-\l_1,\l_1)}{-i\l_1}\right]\\
&=\sqrt{2}\frac{c^2}{\l_1^2}.
\end{split}
\ee
\subsection{Multi-particle states}
Having the above explicit calculation in mind. Now we turn to the multi-particle case. It's obvious that we have the following relation
\be
\begin{split}
&\int_0^{L}dx_N\int_0^{x_{N}}dx_{N-1}\cdots\int_0^{x_2}dx_1A(x_1,\cdots,x_N|\l_1,\cdots,\l_N)\\
&=B_N(\l_1,\cdots,\l_N;a_1,\cdots,a_N)\prod_{j<k}f(\l_j,\l_k),
\end{split}
\ee
where we have defined the basic integral in the region $0<x_1<x_2<\cdots<x_N<L$ as before
\be
B_N(\l_1,\cdots,\l_N;a_1,\cdots,a_N)=\int_0^{L}dx_N\int_0^{x_{N}}dx_{N-1}\cdots\int_0^{x_2}dx_1e^{i(\l_1x_1+\l_2x_2\cdots+\l_Nx_N)}.
\ee
\par For lower $N$, $B_N$ can be integrated out directly. For example
\be\label{B3int}
B_3=\frac{a_1a_2a_3-1}{i\l_1i(\l_1+\l_2)i(\l_1+\l_2+\l_3)}-\frac{a_2a_3-1}{i\l_1i\l_2i(\l_2+\l_3)}+
\frac{a_3-1}{i\l_2i\l_3i(\l_1+\l_2)}.
\ee
For general $N$, it's not easy to directly work out this integral. However, we have the following recursion relation
\be\label{rec}
\frac{\partial B_N(\bm{\l}_N;\bm{a}_N)}{\partial L}=a_NB_{N-1}(\bm{\l}_{N-1};\bm{a}_{N-1}),
\ee
where we view $B_N(\bm{\l}_N;\bm{a}_N)$ as a function of the system size $L$ inexplicitly as the $L$-dependence is encoded in the definition of the variables $a_i$ through eq.~(\ref{a}).
We also have the initial conditions for each $N$
\be\label{bd}
B_N(\bm{\l}_N;\bm{a}_N)|_{L=0}=0, \qquad N=1,2,3,\cdots.
\ee
For example, from
\be
B_1(\l_1;a_1)=\frac{a_1-1}{i\l_1},
\ee
we can get $B_2$ as
\be
\begin{split}
B_2(\l_1,\l_2;a_1,a_2)&=\int_0^La_2B_1(\l_1;a_1)|_{L\rightarrow L'}dL'
=\frac{1}{i\l_1}\int_0^L[e^{i(\l_1+\l_2)L'}-e^{i\l_2L'}]dL'\\
&=\frac{e^{i(\l_1+\l_2)L-1}}{i\l_1i(\l_1+\l_2)}-\frac{e^{i\l_2L}-1}{i\l_1i\l_2}
=\frac{a_1a_2-1}{i\l_1i(\l_1+\l_2)}-\frac{a_2-1}{i\l_1i\l_2},
\end{split}
\ee
which gives the same result with eq.~(\ref{B2int}). In a similar way, we can reproduce the result of eq.~(\ref{B3int}).
\par In the above calculation, it not hard to figure out the pattern. Using the recursion relation eq.~(\ref{rec}) and the initial condition eq.~(\ref{rec}) repeatedly, we obtain the general expressions for $B_N$.
\be\label{B1}
B_N(\bm{\l}_N;\bm{a}_N)=\sum_{j=1}^N(-1)^{j+1}\frac{\prod_{k=j}^Na_k-1}{(\prod_{k=j}^N\sum_{m=j}^k(i\l_m))(\prod_{k=1}^{j-1}\sum_{m=k}^{j-1}(i\l_m))}
\ee
or
\be\label{B2}
\begin{split}
B_N(\bm{\l}_N;\bm{a}_N)&=\sum_{j=0}^N(-1)^{j}\frac{\prod_{k=j+1}^Na_k}{(\prod_{k=j+1}^N\sum_{m=j+1}^k(i\l_m))(\prod_{k=1}^{j}\sum_{m=k}^{j}(i\l_m))}\\
&=\sum_{j=0}^NB_{N,j}(\bm{\l}_N;\bm{a}_N),
\end{split}
\ee
where we define
\be
B_{N,j}(\bm{\l}_N;\bm{a}_N)=(-1)^{j}\frac{\prod_{k=j+1}^Na_k}{(\prod_{k=j+1}^N\sum_{m=j+1}^k(i\l_m))(\prod_{k=1}^{j}\sum_{m=k}^{j}(i\l_m))}.
\ee
These two formulas for $B_N$ in eq.~(\ref{B1}) and eq.~(\ref{B2}) have slight differences but are equivalent to each other. However, we found the later is more useful for our convenience.
\par The $N$-particle formal overlap can be written as
\be
\begin{split}
&\braket{BEC|\bm{\l}_N}=\int d^N\bm{x}_N\braket{BEC|\bm{x}_N}\braket{\bm{x}_N|\bm{\l}_N}\\
&=\frac{c^{N/2}}{L^{N/2}\sqrt{N!}}\sum_{\s\in S_N}\int d^N\bm{x}_NA(\bm{x}_N|\s\bm{\l}_N)\exp\{i\sum_{k=1}^N\l_{\s_k}x_k\}\\
&=\frac{c^{N/2}N!}{L^{N/2}\sqrt{N!}}\int d^N\bm{x}_NA(\bm{x}_N|\bm{\l}_N)\exp\{i\sum_{k=1}^N\l_{k}x_k\}\\
&=\sqrt{\frac{N!c^N}{L^N}}\sum_{\s\in S_N}B_N(\l_{\s_1},\cdots\l_{\s_N};a_{\s_1},\cdots,a_{\s_N})\prod_{j<k}f(\l_{\s_j},\l_{\s_k})\\
&=\sqrt{\frac{N!c^N}{L^N}}\mathcal{S}_N(\bm{\l}_N;\bm{a}_N).
\end{split}
\ee
In the above expression, up to some constant factor, we have found the expression for N-particle formal overlap
\be\label{S}
\mathcal{S}_N(\bm{\l}_N;\bm{a}_N)=\sum_{\s\in S_N}B_N(\l_{\s_1},\cdots\l_{\s_N};a_{\s_1},\cdots,a_{\s_N})\prod_{j<k}f(\l_{\s_j},\l_{\s_k}).
\ee
In \cite{de2014solution}, it was argued that the BEC state is an integrable state. Hence for generic (non-parity-symmetric) on-shell Bethe state, $\mathcal{S}_N(\bm{\l}_N;\bm{a}_N)$ vanishes. Only for the parity symmetric on-shell states $\ket{\bm{\l}_{N/2}^+,-\bm{\l}_{N/2}^+}$, we have finite overlaps. See Appendix \ref{appenA} for a direct proof of this statement.
\subsubsection{Determining the singular part}
In order to find out the non-vanishing overlap of on-shell parity symmetric Bethe states with the BEC state, we only need to figure out the singular part of $S_N(\bm{\l}_N;\bm{a}_N)$ and taking the limiting procedure.
\par Firstly, we must try to find out the residue
\be
\mathrm{Res}_{\l_{m+1}\rightarrow-\l_m}B_N(\bm{\l}_N;\bm{a}_N).
\ee
It's easy to find that there are two terms $B_{N,m-1}$ and $B_{N,m+1}$ contribute
\be
\begin{split}
&\frac{(-1)^{m-1}\prod_{k=m}^Na_k}{(\prod_{k=m}^N\sum_{l=m}^k(i\l_l))(\prod_{k=1}^{m-1}\sum_{l=k}^{m-1}(i\l_l))}
+\frac{(-1)^{m+1}\prod_{k=m+2}^Na_k}{(\prod_{k=m+2}^N\sum_{l=m+2}^k(i\l_l))(\prod_{k=1}^{m+1}\sum_{l=k}^{m+1}(i\l_l))}\\
&=\frac{(-1)^{m-1}a_ma_{m+1}\prod_{k=m+2}^Na_k}{i\l_mi(\l_m+\l_{m+1})(\prod_{k=m+2}^N\sum_{l=m+2}^k(i\l_l))(\prod_{k=1}^{m-1}\sum_{l=k}^{m-1}(i\l_l))}\\
&+\frac{(-1)^{m-1}\prod_{k=m+2}^Na_k}{i\l_{m+1}i(\l_m+\l_{m+1})(\prod_{k=m+2}^N\sum_{l=m+2}^k(i\l_l))(\prod_{k=1}^{m-1}\sum_{l=k}^{m-1}(i\l_l))}.
\end{split}
\ee
Then the singular part of $B_N$ is
\be\label{singular}
B_N(\bm{\l}_N;\bm{a}_N)\sim \frac{a_ma_{m+1}-1}{i(\l_{m}+\l_{m+1})}\frac{1}{i\l_m}
\frac{(-1)^{m-1}\prod_{k=m+2}^Na_k}{(\prod_{k=m+2}^N\sum_{l=m+2}^k(i\l_l))(\prod_{k=1}^{m-1}\sum_{l=k}^{m-1}(i\l_l))}.
\ee
The relation eq.~(\ref{singular}) can be rewritten as
\be
\begin{split}
B_N(\{\l_j\};\{a_j\}|j\in\{1,\cdots,N\})\sim \frac{a_ma_{m+1}-1}{i(\l_{m}+\l_{m+1})}\frac{1}{i\l_m}\\
\times B_{N-2,m-1}(\{\l_j\};\{a_j\}|j\in\{1,\cdots,\cancel{m},\cancel{m+1},\cdots,N\}).
\end{split}
\ee
\par After the summation of all permutations that make particles $m$ and $m+1$ in neighbouring position, we find the singularity of the formal overlap at $\l_{m+1}=-\l_m$ is
\be\label{Srec}
\mathcal{S}_N\sim\frac{a_ma_{m+1}-1}{i(\l_m+\l_{m+1})}F(\l_m)\prod_{\substack{j=1\\j\neq m,m+1}}^Nf(\l_j,\l_m)f(\l_j,-\l_m)\mathcal{S}_{N-2}^{\mathrm{mod}}(\cancel{m},\cancel{m+1}),
\ee
where $\mathcal{S}_{N-2}^{\mathrm{mod}}(\cancel{m},\cancel{m+1})$ is the formal overlap for $N-2$ particles which does not include particles $m$ and $m+1$, and is evaluated with the modified a-variables\footnote{Here, the definition of the modified a-variables is slightly different from the definition given in \cite{Jiang:2020sdw}.}:
\be
a_j^{\mathrm{mod}}=\frac{f(\l_m,\l_j)}{f(\l_j,\l_m)}\frac{f(-\l_m,\l_j)}{f(\l_j,-\l_m)}a_j.
\ee
The function $F(\l)$ can be found as
\be\label{F}
F(\l)=\frac{f(\l,-\l)}{i\l}+\frac{f(-\l,\l)}{-i\l}=\frac{c}{\l^2}.
\ee
\subsubsection{Taking the limiting procedure}
The exact on-shell parity symmetric Bethe state with the BEC state overlap is obtained only after the following limiting procedure is token
\be\label{limit}
\l_{2j-1}\rightarrow \l_j^+,\qquad \l_{2j}\rightarrow -\l_j^+,\qquad\quad j=1,2,,\cdots,N.
\ee
Note that
\be
\begin{split}
&\lim_{\substack{\l_{2j-1}\rightarrow\l_j^+ \\\l_{2j}\rightarrow-\l_j^+}}\frac{a_{2j-1}a_{2j}-1}{i(\l_{2j-1}+\l_{2j})}=\lim_{\l\rightarrow\l_{j}^+}\frac{a(\l)a(-\l_{j}^+)-1}{i(\l-\l_{j}^+)}\\
&=\lim_{\l\rightarrow\l_{j}^+}\frac{a'(\l)a(-\l_j^+)a(\l)}{ia(\l)}=-i\frac{d}{d\l}\log(a(\l))\Bigg|_{\l=\l_j^+}.
\end{split}
\ee
Then it's useful to introduce new variables
\be
m(\l)=-i\frac{d}{d\l}\log(a(\l)), \qquad m_j^+\equiv m(\l_j^+),\quad  j=1,2,\cdots,\frac{N}{2}.
\ee
We denote $D(\bm{\l}_{N/2}^+,\bm{m}_{N/2}^+)$ as the limit of $\mathcal{S}_N(\bm{\l}_N;\bm{a}_N)$ describe in eq.~(\ref{limit}).

It's easy to show that the function $D(\bm{\l}_{N/2}^+,\bm{m}_{N/2}^+)$ satisfies the following recursion relation
\be\label{rec1}
\frac{\partial D(\bm{\l}_{N/2}^+,\bm{m}_{N/2}^+)}{\partial m_j^+}=F(\l_j^+)\prod_{k=1,k\neq j}^{N/2}\bar{f}(\l_k^+,\l_j^+)D(\bm{\l}_{N/2-1}^+,\bm{m}_{N/2-1}^{+,\mathrm{mod}}),
\ee
where the modification rule for the m-parameters is
\be
m^{\mathrm{mod}}(\l)=-i\frac{d}{d\l}\log(a^{\mathrm{mod}}(\l))=m(\l)+\varphi^+(\l,\l_j^+),
\ee
and $D(\bm{\l}_{N/2-1}^+,\bm{m}_{N/2-1}^{+,\mathrm{mod}})$ is understood as the $j$-th variables have removed from it's arguments.
The solution of the recursion relation eq.~(\ref{rec1}) can be found as \cite{Jiang:2020sdw}
\be
D(\bm{\l}_{N/2}^+,\bm{m}_{N/2}^+)=\prod_{j=1}^{N/2}F(\l_j^+)\prod_{1\leq k<j\leq N/2}\bar{f}(\l_k^+,\l_j^+) \det G^+.
\ee
\par Then the exact on-shell parity symmetric Bethe state with the BEC state overlap is
\be\label{Overlap1}
\frac{|\braket{BEC|\bm{\l}_{N/2}^+,-\bm{\l}_{N/2}^+}|}{\parallel\ket{\bm{\l}_{N/2}^+,-\bm{\l}_{N/2}^+}\parallel}=\sqrt{\frac{N!}{L^N}}
\prod_{j=1}^{N/2}\frac{F(\l_j^+)}{\sqrt{f(\l_j^+,-\l_j^+)f(-\l_j^+,\l_j^+)}}\sqrt{\frac{\det G^+}{\det G^-}}.
\ee
After the substitution of the expression eq.~(\ref{F}) and eq.~(\ref{f}) into eq.~(\ref{Overlap1}), we obtained the final result
\be
\frac{|\braket{BEC|\bm{\l}_{N/2}^+,-\bm{\l}_{N/2}^+}|}{\parallel\ket{\bm{\l}_{N/2}^+,-\bm{\l}_{N/2}^+}\parallel}=\frac{\sqrt{N!(cL)^{-N}}}{\prod_{j=1}^{N/2}\frac{\l_j^+}{c}\sqrt{\frac{\l_j^{+2}}{c^2}+\frac14}}
\sqrt{\frac{\det G^+}{\det G^-}}.
\ee
This formula agrees with the earlier results \cite{de2014solution,brockmann2014overlaps}.
\section{Overlap between the Bethe state and free particle state with pair structure}\label{sec4}
As an application of our formula eq.~(\ref{B2}) obtained in the last section, in this section we consider the overlap between the Bethe state and a special kind of free particle states $\ket{\Psi_{\bm{k}_N}}$. The momenta of the particles are paired up leading to zero total momentum of these states. The wave function of $\ket{\Psi_{\bm{k}_N}}$ can be written as
\be\label{Psi}
\braket{\bm{x}_N|\Psi_{\bm{k}_N}}=C_N\sum_{\s\in S_N}e^{i\bm{k}_N\cdot\s\bm{x}_N}=C_N\sum_{\s\in S_N}e^{i\s\bm{k}_N\cdot\bm{x}_N}=C_N\sum_{\s\in S_N}e^{i\sum_{j=1}^Nk_{\s_j}x_j},
\ee
where
\be
\{\bm{k}_N\}=\{k,-k,\cdots,k,-k\}\equiv\{\bm{k}_{N/2},-\bm{k}_{N/2}\},
\ee
and $C_N$ is the normalization constant. In the following part, we will denote $\ket{\Psi_{\bm{k}_N}}$ as $\ket{\Psi_{\bm{k}_{N/2},-\bm{k}_{N/2}}}$ to emphasize its pair structure.
\par Let's begin by two-particle state
\be\label{Psi2}
\braket{x_1,x_2|\Psi_{k,-k}}=C_2(e^{ikx_1-ikx_2}+e^{ikx_2-ikx_1}).
\ee
The normalization constant is $k$-dependent: $C_2=1/\sqrt{2(L^2+\frac{\sin^2(kL)}{k^2})}$. When $\ket{\Psi_{\bm{k}_{N/2},-\bm{k}_{N/2}}}$  is on-shell, \textup{i.e.} we impose periodic condition to these wave function eq.~(\ref{Psi}), which means $e^{ikL}=1$. Then $C_2=1/\sqrt{2L^2}$.
\par Then the two-particle overlap can be compute directly
\be\label{fS2}
\begin{split}
&\braket{\Psi_{k,-k}|\l_1,\l_2}=\frac{2c}{\sqrt{2!}\sqrt{2L^2}}\Big\{f(\l_1,\l_2)\int_0^Ldx_2\int_0^{x_2}dx_1(e^{i(\l_1+k)x_1+i(\l_2-k)x_2}+e^{i(\l_1-k)x_1+i(\l_2+k)x_2})\\
&+f(\l_2,\l_1)\int_0^Ldx_1\int_0^{x_1}dx_2(e^{i(\l_1+k)x_1+i(\l_2-k)x_2}+e^{i(\l_1-k)x_1+i(\l_2+k)x_2})\Big\}\\
&=\frac{c}{L}\{f(\l_1,\l_2)[B_2(\l_1+k,\l_2-k;a_1,a_2)+B_2(\l_1-k,\l_2+k;a_1,a_2)]+(\l_1\leftrightarrow\l_2,a_1\leftrightarrow a_2)\},
\end{split}
\ee
where we have assumed $\ket{\Psi_{k,-k}}$ is on-shell. It's then easily to check that when the Bethe state is also on-shell, $\textup{i.e}$ when we substitute the Bethe equations
eq.~(\ref{Bethe}) into eq.~(\ref{fS2}), we obtain zero.  This implies that $\ket{\Psi_{k,-k}}$ is an integrable state\footnote{It's easy to check that $Q_3\ket{\Psi_{k,-k}}=0$}.
\par The non-vanishing overlap can be obtained only for Bethe states with parity symmetry as before. We only need to figure out the singular pieces in the formal overlap eq.~(\ref{fS2}) and taking the limiting procedure. Around the pole $\l_2=-\l_1$, we have
\be
\begin{split}
\braket{\Psi_{k,-k}|\l_1,\l_2}&\sim \frac{a_1a_2-1}{i(\l_1+\l_2)}\frac{c}{L}\Big\{f(\l_1,\l_2)\left[\frac{1}{i(\l_1+k)}+\frac{1}{i(\l_1-k)}\right]\\
&+f(\l_2,\l_1)\left[\frac{1}{i(\l_2+k)}+\frac{1}{i(\l_2-k)}\right]\Big\}\\
&=\frac{a_1a_2-1}{i(\l_1+\l_2)}\frac{c}{L}\left[f(\l_1,\l_2)\frac{-2i\l_1}{\l_1^2-k^2}+f(\l_2,\l_1)\frac{-2i\l_2}{\l_2^2-k^2}\right].
\end{split}
\ee
Then the two-particle overlap is
\be
\begin{split}
\braket{\Psi_{k,-k}|\l_1,-\l_1}&=\lim_{\l_2\rightarrow -\l_1}\frac{e^{i(\l_1+\l_2)L}-1}{i(\l_1+\l_2)}\frac{c}{L}\left[f(\l_1,\l_2)\frac{-2i\l_1}{\l_1^2-k^2}+f(\l_2,\l_1)\frac{-2i\l_2}{\l_2^2-k^2}\right]\\
&=c\left[f(\l_1,-\l_1)\frac{-2i\l_1}{\l_1^2-k^2}+f(-\l_1,\l_1)\frac{2i\l_1}{\l_1^2-k^2}\right]\\
&=\frac{2c^2}{\l_1^2-k^2}.
\end{split}
\ee
\par For the general $N$-particle states, the normalization constant can be fixed as $C_N=\frac{1}{(N/2)!\sqrt{N!}L^{N/2}}$ when $\ket{\Psi_{\bm{k}_{N/2},-\bm{k}_{N/2}}}$ is on-shell. We can compute the N-particle formal overlap as
\be\label{fS0}
\begin{split}
&\braket{\Psi_{\bm{k}_{N/2},-\bm{k}_{N/2}}|\bm{\l}_N}=\int d^N\bm{x}_N\braket{\Psi_{\bm{k}_{N/2},-\bm{k}_{N/2}}|\bm{x}_N}\braket{\bm{x}_N|\bm{\l}_N}\\
&=\frac{c^{N/2}}{(N/2)!L^{N/2}\sqrt{N!}\sqrt{N!}}\sum_{\s\in S_N}\int d^N\bm{x}_NA(\bm{x}_N|\s\bm{\l}_N)\exp\{i\sum_{k=1}^N\l_{\s_k}x_k\}\sum_{\t\in S_N}e^{i\sum_{j=1}^Nk_{\t_j}x_j}\\
&=\frac{c^{N/2}N!}{(N/2)!L^{N/2}\sqrt{N!}\sqrt{N!}}\int d^N\bm{x}_NA(\bm{x}_N|\bm{\l}_N)\exp\{i\sum_{k=1}^N\l_{k}x_k\}\sum_{\t\in S_N}e^{i\sum_{j=1}^Nk_{\t_j}x_j}\\
&=\frac{c^{N/2}}{(N/2)!L^{N/2}}\int d^N\bm{x}_N\sum_{\tau\in S_N}e^{i(\bm{\l}_N+\t\bm{k}_N)\cdot\bm{x}_N}A(\bm{x}_N|\bm{\l}_N)\\
&=\frac{c^{N/2}}{(N/2)!L^{N/2}}\sum_{\s\in S_N}\mathcal{B}_N(\s\bm{\l}_N;\s\bm{a}_N)\prod_{j<k}f(\l_{\s_j},\l_{\s_k})\\
&=\frac{c^{N/2}}{(N/2)!L^{N/2}}\mathfrak{S}_N(\bm{\l}_N;\bm{a}_N),
\end{split}
\ee
where
\be
\mathcal{B}_N(\bm{\l}_N;\bm{a}_N)=\sum_{j=0}^N\mathcal{B}_{N,j}(\bm{\l}_N;\bm{a}_N)
\ee
with
\be
\mathcal{B}_{N,j}(\bm{\l}_N;\bm{a}_N)=\sum_{\t\in S_N}B_{N,j}(\bm{\l}_N+\t\bm{k}_N;\bm{a}_N).
\ee
In the above expression, up to some constant factor, we have found that the N-particle formal overlap can be expressed as
\be\label{fS}
\mathfrak{S}_N(\bm{\l}_N;\bm{a}_N)=\sum_{\s\in S_N}\mathcal{B}_N(\s\bm{\l}_N;\s\bm{a}_N)\prod_{j<k}f(\l_{\s_j},\l_{\s_k}).
\ee
\par Unfortunately, for general $N$ ($N\geq4$), since $Q_3\ket{\Psi_{\bm{k}_{N/2},-\bm{k}_{N/2}}}\neq 0$, the states become non-integrable and they can have non-zero overlaps with non-parity-symmetric Bethe states. We still consider the overlaps with the parity symmetric Bethe states for simplicity. In this case, the strategy to obtain exact overlaps described in the last section becomes invalid. However, since the singular part in the formal overlap eq.~(\ref{fS}) dominant in the large $L$ limit \cite{deLeeuw:2019sew}, we can still rely on previous analysis to obtain the overlaps in the large $L$ limit.
\par The singular part of $\mathcal{B}_N$ at the pole $\l_{m+1}=-\l_m$ can be extract from $\mathcal{B}_{N,m-1}$ and $\mathcal{B}_{N,m+1}$. Each contains a summation of permutation of $\bm{k}_{N}$ in the argument of the corresponding $B_{N,j}$. Only the permutations that satisfy $k_{\tau_{m}}=k,k_{\tau_{m+1}}=-k$ or $k_{\tau_{m}}=-k,k_{\tau_{m+1}}=k$ contribute. Then the calculation is similar to that was presented in the last section. We find the following result
\be
\begin{split}
&\mathcal{B}_N(\{\l_j\};\{a_j\}|j\in\{1,\cdots,N\})\sim \frac{N^2(N-2)^2}{2^4}\frac{a_ma_{m+1}-1}{i(\l_{m}+\l_{m+1})}\left[\frac{1}{i(\l_m+k)}+\frac{1}{i(\l_m-k)}\right]\\
&\times \mathcal{B}_{N-2,m-1}(\{\l_j\};\{a_j\}|j\in\{1,\cdots,\cancel{m},\cancel{m+1},\cdots,N\})\\
&=\frac{N^2(N-2)^2}{2^4}\frac{a_ma_{m+1}-1}{i(\l_{m}+\l_{m+1})}\frac{2\l_m}{i(\l_m^2-k^2)}\mathcal{B}_{N-2,m-1}(\{\l_j\};\{a_j\}|j\in\{1,\cdots,\cancel{m},\cancel{m+1},\cdots,N\}).
\end{split}
\ee
\par After the summation of all permutations that make particles $m$ and $m+1$ in neighbouring position, we find the singularity of the formal overlap $\mathfrak{S}_N$ at $\l_{m+1}=-\l_m$ essentially has the same recursion relation with eq.~(\ref{Srec})
\be
\mathfrak{S}_N\sim\frac{N^2(N-2)^2}{2^4}\frac{a_ma_{m+1}-1}{i(\l_m+\l_{m+1})}F(\l_m)\prod_{\substack{j=1\\j\neq m,m+1}}^Nf(\l_j,\l_m)f(\l_j,-\l_m)\mathfrak{S}_{N-2}^{\mathrm{mod}}(\cancel{m},\cancel{m+1}),
\ee
but in this case
\be
F(\l)=\frac{2\l}{i(\l^2-k^2)}f(\l,-\l)+\frac{-2\l}{i(\l^2-k^2)}f(-\l,\l)=\frac{2c}{\l^2-k^2}.
\ee
\par Then applying the same discussion in the last section and noting that the ratio of the determinant $\sqrt{\frac{\det G^+}{\det G^-}}$ only gives $\mathcal{O}(1)$ pieces in the large $L$ limit, we arrive at our final result
\be
\frac{|\braket{\Psi_{\bm{k}_{N/2},-\bm{k}_{N/2}}|\bm{\l}_{N/2}^+,-\bm{\l}_{N/2}^+}|}{\parallel\ket{\bm{\l}_{N/2}^+,-\bm{\l}_{N/2}^+}\parallel}=\frac{(N/2)!}{L^{N/2}}\left[\prod_{j=1}^{N/2}\frac{4c}{
(\l_j^{+2}-k^2)\sqrt{\frac{c^2}{\l^{+2}_j}+4}}+\mathcal{O}(L^{-1})\right].
\ee
The same strategy presented in this section can be applied to calculating overlaps of the Bethe states with other initial states with simple wave functions in the large $L$ limit \cite{bucciantini2016stationary}.
\section{Conclusion}\label{sec5}
In this note, we derived the exact overlap formula between the Lieb-Liniger Bethe states and the Bose-Einstein condensate state using a recently developed method. This method is based on the coordinate Bethe ansatz which is available for the Lieb-Liniger model and does not relies on the complicated ``rotation trick'' which is not known in our case. This overlap formula is of great importance in the study of the BEC to Lieb-Liniger quench, but a transparent derivation is lacking. This paper gives a rather concise computation procedure based on the newly proposed method. We also obtained overlap for a special kind of free particle states in the large $L$ limit. This formula serves as a starting point to the study of the related quench problem.\
\par We hope more insight can be found in this new method and it's very interesting to find more applications such as the nest integrable system where similar exact overlap formulae also exist (see for example \cite{deLeeuw:2016umh,deLeeuw:2018mkd,deLeeuw:2019ebw}).
\par In integrable spin chain models, there also exist exact overlap formulae for integrable Matrix Product States (MPS) \cite{Buhl-Mortensen:2015gfd,deLeeuw:2018mkd,Pozsgay:2018dzs}. In \cite{maruyama2010continuous} the authors introduced the continuous Matrix Product States (cMPS) in the Lieb-Liniger model. It would be very interesting to find out exact overlap formulae for integrable cMPS.\
\par As you will see in the appendix, the proof of the overlap vanishes for generic (non-parity-symmetric) on-shell Bethe is very specific and is not inspiring (for an earlier proof, see \cite{brockmann2014neel}). A more transparent proof is needed.\
\par It turns out that the coordinate Bethe ansatz and the calculation of exact overlap have some applications in the physics of the stochastic non-linear Kardar-Parisi-Zhang (KPZ) equation, where the computation of the exact generating function of the Kardar-Parisi-Zhang height requires to calculate the overlap between the Bethe wave functions and the initial condition of the equation \cite{calabrese2011exact,calabrese2014interaction}. We think our result and all the classification obtained in \cite{Piroli:2017sei} could help to classify the integrable initial conditions in the KPZ equation.
\section*{Acknowledgments}
I would like to thank  Hao Ouyang for very helpful discussions, especially for the proof of the vanishing of the overlap for generic on-shell Bethe states with the BEC state.
\begin{appendix}
\section{Proof of $\mathcal{S}_N=0$ for generic on-shell Bethe state}\label{appenA}
In this appendix, let's prove that the overlap of a generic on-shell Bethe state with the BEC state vanishes.
\par For a given $\sigma$, let's consider terms in eq.~(\ref{S}) containing the factor
\begin{equation}
F_\s\equiv\prod_{j<k}f(\l_{\s_{j}},\l_{\s_{k}}).
\end{equation}
If we define $\l_{\s_{j}}=i p^\s_{{j}}-i p^\s_{{j-1}}$, then we have
\begin{equation}
F_{\s}B_{N,j}(\l_{\s_{1}},...,\l_{\s_{N}};a_{\s_1},...,a_{\s_N})
=F_{\tau_j\s}\prod^N_{\substack{k=0 \\ k\neq j}}\frac{1}{p^\s_{{j}}-p^\s_{{k}}},
\end{equation}
where $\tau_j\in S_N$ and is defined as
\begin{equation}
\tau_j=\left(\begin{array}{c}
1,...,j,j+1,...,N\\
j+1,...,N,1,...,j
\end{array} \right),
\end{equation}
when $0<j<N$, and $\tau_0$ and $\tau_N$ are identity.
Then, for a generic on-shell Bethe state, we have
\be\label{onS}
\begin{split}
\mathcal{S}_N(\bm{\l}_N;\bm{a}_N)=\sum_{\s\in S_N}\sum_{j=0}^NF_{\tau_j\s}\prod^N_{\substack{k=0 \\ k\neq j}}\frac{1}{p^\s_{{j}}-p^\s_{{k}}}\\
=\sum_{\s\in S_N}F_{\s}\sum_{j=0}^N\prod^N_{\substack{k=0 \\ k\neq j}}\frac{1}{p^{\tau_j^{-1}\s}_{{j}}-p^{\tau_j^{-1}\s}_{k}}
\end{split}
\ee
It is not difficult to see that the coefficient of the $F_\s$ term in eq.~(\ref{onS}) vanish. In fact, for any permutation $\s\in S_N$, we have
\begin{equation}\label{coeff}
\sum_{j=0}^{N}\prod^N_{\substack{k=0 \\ k\neq j}}\frac{1}{p^{\tau_j^{-1}\s}_{{j}}-p^{\tau_j^{-1}\s}_{k}}=0.
\end{equation}
The terms in eq.~(\ref{coeff}) with $j=0$ and $j=N$ come from the summand in eq.~(\ref{S}) corresponding to the permutation $\sigma$.
 and the term with $0<j<N$ comes from the summand in eq.~(\ref{S}) corresponding to the permutation $\tau_j^{-1}\s$.
\end{appendix}
\\


\bibliography{2020}
\bibliographystyle{ieeetr}


\end{document}